\documentclass[12pt]{article}
\pdfoutput=1
\usepackage{textcomp}
\usepackage{color}
\usepackage{putex}
\usepackage{autobreak}
\usepackage{indentfirst}
\usepackage{graphicx}
\usepackage{float}
\graphicspath{{plots/}}
\usepackage{tabularx}
\usepackage{caption}
\usepackage{amsmath}
\numberwithin{equation}{section}
\usepackage{cancel}
\usepackage{array}
\usepackage{subcaption}
\usepackage{epstopdf}
\usepackage{enumerate}
\usepackage{cite}
\usepackage{tensor}
\usepackage{youngtab}
\usepackage{tensor}
\usepackage{slashed}
\usepackage[aligntableaux=center]{ytableau}
\usepackage{CJKutf8}
\usepackage[utf8]{inputenc}
\usepackage{rotating}
\usepackage{multirow}
\usepackage[
      colorlinks=true,
      linkcolor=red,
      urlcolor=red,
      filecolor=black,
      citecolor=red,
      ]{hyperref}
\usepackage[noabbrev]{cleveref}
\usepackage{tikz, pgfplots}
\usetikzlibrary{arrows.meta, calc, positioning}
\usetikzlibrary{decorations.pathmorphing}
\tikzset
{
    node/.style=
    {
        draw=black, 
        fill=white, 
        circle, 
        minimum width=2cm, 
        inner sep=45pt
    },
    blacknode/.style=
    {
    circle, 
    inner sep=3pt, 
    fill=black
    },
    rednode/.style=
    {
    circle, 
    inner sep=7pt, 
    fill=red
    }
}
\begin{document}
\institution{forbiddencity}{Kavli Institute for Theoretical Sciences, University of Chinese Academy of Sciences, \cr Beijing 100190, China.}
\institution{summerpalace}{School of Physical Sciences, University of Chinese Academy of Sciences, No.19A Yuquan Road, \cr Beijing 100049, China.}
\title{
Bootstrapping AdS$_2 \times$ S$^2$ hypermultiplets: hidden four-dimensional conformal symmetry
}
\authors{
Konstantinos C. Rigatos\worksat{\forbiddencity}${}^{,}$\footnote{{\hypersetup{urlcolor=black}\href{mailto:rkc@ucas.ac.cn}{rkc@ucas.ac.cn}}}
and
Shaodong Zhou\worksat{\forbiddencity}${}^{,}$\worksat{\summerpalace}${}^{,}$\footnote{{\hypersetup{urlcolor=black}\href{mailto:zhoushaodong17@mails.ucas.ac.cn}{zhoushaodong17@mails.ucas.ac.cn}}}
}
\abstract{
We bootstrap the $4$-point amplitude of $\mathcal{N}=2$ hypermultiplets in $\text{AdS}_2 \times \text{S}^2$ at tree-level and for arbitrary external weights. We hereby explicitly demonstrate the existence of a hidden four-dimensional conformal symmetry that was used as an assumption in previous studies to derive this result.
}
\date{\today}
\maketitle
{
\hypersetup{linkcolor=black}
\tableofcontents
}
\newpage
\section{Prologue}\label{sec: prologue}
Correlation functions of local operators are the most basic and natural observables to study in any (super)conformal field theory. By virtue of the AdS/CFT duality they are dual to on-shell scattering amplitudes in AdS and in the holographic limit these observables are expanded in powers of the inverse central charge. To leading order holographic correlators are given just by generalized free field theory. To extract non-trivial dynamical information we need to consider higher orders in the central charge expansion. The computation of sub-leading contributions is burdensome from the CFT side owing to the theory being strongly coupled. In the weakly coupled dual description it is possible to perform these calculations, at least in principle. Traditionally one would have to resort to a diagrammatic expansion in AdS. It should be noted, however, that this approach requires the precise knowledge of the effective Lagrangians and due to the proliferation of diagrams and complicated vertices, see for instance \cite{Arutyunov:1999fb}, it has been rather impractical to use and results were obtained in the early days for a handful of examples \cite{Freedman:1998tz,DHoker:1999kzh,Arutyunov:2000py,Arutyunov:2002fh,Arutyunov:2003ae}. Furthermore, while this approach is conceptually straightforward, the computations become quickly unwieldy and hence the form of the answer lacks any suggestive structure.

It was only in recent years that we have understood a truly effective approach to compute these holographic correlators and since then we have witnessed a profusion of significant results in these studies in different regimes of the expansion. These new developments are based on a different strategy altogether. In this modern approach, we work directly with the holographic correlators and use superconformal symmetry and other consistency conditions to fix the result. One of the upshots of this bootstrap approach is that it shuns the need of an explicit effective Lagrangian. This method was initiated in \cite{Rastelli:2016nze,Rastelli:2017udc} and led to the complete $4$-point functions of $\tfrac{1}{2}$-BPS operators with arbitrary Kaluza-Klein (KK) levels at tree-level\footnote{Highly non-trivial checks have been performed in \cite{Arutyunov:2017dti,Arutyunov:2018tvn,Arutyunov:2018neq} by explicit supergravity computations.}. This paved the way for an array of very impressive results. From the tree-level correlators we can extract the CFT data for unprotected double-trace operators \cite{Aprile:2017bgs,Aprile:2017xsp,Aprile:2018efk,Alday:2017xua}. In turn, we can proceed by considering these as input to obtain results at one-loop \cite{Aprile:2017qoy,Alday:2017vkk,Alday:2019nin,Aprile:2019rep}, subsequently move on to two-loops \cite{Huang:2021xws,Drummond:2022dxw}, and even use this bootstrap approach to extend the studies beyond the $4$-point case \cite{Goncalves:2019znr,Goncalves:2023oyx}. Not only that, but one can consider stringy corrections to the $4$-point correlators \cite{Goncalves:2014ffa,Alday:2018pdi,Binder:2019jwn,Drummond:2019odu,Drummond:2020uni,Drummond:2020dwr,Aprile:2020mus,Aprile:2022tzr,Drummond:2019hel}.

This very beautiful story has been unfolding to different extents in other backgrounds as well. The techniques developed in $\text{AdS}_5 \times \text{S}^5$ have been used to provide us with a plethora of results in $\text{AdS}_7 \times \text{S}^4$ \cite{Rastelli:2017ymc,Alday:2020lbp,Zhou:2017zaw,Heslop:2017sco,Abl:2019jhh,Alday:2020tgi} and $\text{AdS}_4 \times \text{S}^7$ \cite{Zhou:2017zaw,Chester:2018lbz,Chester:2018aca,Binder:2018yvd,Alday:2021ymb,Alday:2020dtb,Alday:2022rly} and $\text{AdS}_3 \times \text{S}^3$ \cite{Rastelli:2019gtj,Aprile:2021mvq,Giusto:2018ovt,Giusto:2019pxc,Giusto:2020neo,Giusto:2020mup} supergravities\footnote{It is worthwhile mentioning that there is a unifying approach to treat all holographic correlation functions of arbitrary $\tfrac{1}{2}$-BPS operators in all maximally superconformal field theories \cite{Alday:2020dtb}.}. Moreover, it has provided us with elaborate results for super Yang-Mills theories in AdS \cite{Zhou:2018ofp,Alday:2021odx,Bissi:2022wuh,Huang:2023oxf,Alday:2021ajh,Alday:2022lkk,Huang:2023ppy,Alday:2023kfm,Drummond:2022dxd,Paul:2023zyr,Glew:2023wik,Cao:2023cwa}. For a more detailed exposition and more advancements to these topics we refer the reader to the recent reviews \cite{Bissi:2022mrs,Heslop:2022xgp}. 

Having a panoply of results available allowed the observation of impressive underlying structures in the descriptions of holographic correlators in some specific setups. These structures are hidden in the sense that they are not obvious in any way from the Lagrangian descriptions of the theories. They are interesting not only from a practical point of view allowing one to obtain more compact and suggestive expressions for the correlation functions, but mainly because they are strong indications of new symmetry properties of the bulk theory, hence sharpening our understanding of the theories under examination. These hidden symmetry structures include the Parisi-Sourlas supersymmetry \cite{Behan:2021pzk}, AdS double copy relations \cite{Zhou:2021gnu}, and the emergence of a hidden conformal symmetry. The latter was first observed in the context of $\text{AdS}_5 \times \text{S}^5$ \cite{Caron-Huot:2018kta} and later in $\text{AdS}_3 \times \text{S}^3$ \cite{Rastelli:2019gtj}, and $\text{AdS}_5 \times \text{S}^3$ \cite{Alday:2021odx}.

In this work we are interested in hidden conformal symmetry and more specifically its status in the context of $\text{AdS}_2 \times \text{S}^2$ supergravity. This background arises after a $\text{T}^7$ compactification of M-theory \cite{Lee:1999yu} and a further reduction on the $\text{S}^2$ yields gravity- and hyper-multiplets \cite{Michelson:1999kn,Lee:1999yu,Corley:1999uz}. The current state of affairs for $\text{AdS}_2 \times \text{S}^2$ is the following: in \cite{Abl:2021mxo} the authors used as a working assumption the existence of a four-dimensional conformal symmetry and managed to derive the unmixed anomalous dimensions of the exchanged double-trace operators. Thereupon, the work of \cite{Heslop:2023gzr} provides evidence for this symmetry being present at the loop-order. 

The existence of such a hidden conformal symmetry in a given theory relies heavily on some crucial facts. It is worthwhile stressing that we still lack a formal and thorough understanding of this hidden symmetry, however, we know that its existence simplifies the computations dramatically, and by now we have obtained some intuitive understanding on when to expect that it will be present. Let us briefly review some of the intuition of \cite{Caron-Huot:2018kta} for $\text{AdS}_5 \times \text{S}^5$ and see how these statements can be extended to our case of interest. To begin with, the metric of the $\text{AdS}_5 \times \text{S}^5$ background is conformally equivalent to that of flat space. This is a feature that is common to  $\text{AdS}_5 \times \text{S}^3$ and $\text{AdS}_3 \times \text{S}^3$, however it is not true for  $\text{AdS}_{4,7} \times \text{S}^{7,4}$. Furthermore, the $10$-dimensional flat-space amplitude of the type IIB theory contains the dimensionless factor $G^{10}_N \delta^{16}(Q)$ that is regarded as a dimensionless coupling. Finally, the $10$-dimensional flat-space amplitude of type IIB is conformally invariant and can be considered as the generating function of all KK modes on $\text{AdS}_5 \times \text{S}^5$. The $\text{AdS}_2 \times \text{S}^2$ background draws many similarities with the above. The metric is, in this case as well, that of flat space up to a conformal factor. In addition to that, the $G_N \delta^4(Q)$ factor that enters in the expression of the flat-space amplitude is dimensionless in $4$ dimensions. And finally, the flat-space amplitude is invariant under the action of the generator of conformal transformations. 

The only qualitative difference when comparing to the situation in the $\text{AdS}_5 \times \text{S}^5$ picture is the existence of two different types of multiplets in our set-up. We have the gravity and hypermultiplets and this is close analogy to the $\text{AdS}_3 \times \text{S}^3$ that possesses gravity and tensor multiplets. It was observed in \cite{Rastelli:2019gtj} that the $4$-point function of tensor multiplets enjoys an accidental $6$-dimensional conformal symmetry and there is a comment that with the current results in the literature it appears that this will not be true for the gravity multiplet. This, and also the study of \cite{Abl:2021mxo}, leads us to focus on the hypermultiplets in this work. Another is due to the subtleties that arise in a $1$-dimensional CFT, one of which concerns the lack of a stress tensor \cite{Qiao:2017xif}. 

Extra physics motivation for our work comes from the study of defects in the context of holography. $\text{AdS}_2$ is ubiquitous in the framework of defects and there is much progress to that end with studies of Wilson lines spanning an $\text{AdS}_2$ subsector within the $\text{AdS}_7 \times \text{S}^4$ \cite{Drukker:2020swu}, $\text{AdS}_5 \times \text{S}^5$ \cite{Ferrero:2021bsb,Ferrero:2023znz,Ferrero:2023gnu}, and $\text{AdS}_5 \times \mathbb{RP}^5$ \cite{Giombi:2020kvo} backgrounds. While these setups are not exactly the same as the one we are considering here, they are closely related.

On top of the discussion so far, there is some mathematical motivation as well. More specifically it is interesting to examine how well the position-space bootstrap can work in this simple setup. This is because, while the Mellin-space bootstrap has been very successful in the higher-dimensional cases, in this specific scenario it is not applicable. This is due to the usual problems and complications that arise when trying to define the Mellin transformation in a $1$-dimensional theory; for a thorough analysis of these subtleties see \cite[section 6]{Bissi:2022mrs} and also \cite{Bianchi:2021piu,Ferrero:2019luz} for related progress to that direction. 

On the contrary, the position space bootstrap approach developed in \cite{Rastelli:2016nze,Rastelli:2017udc} can be straightforwardly applied in the case of $\text{AdS}_2$. In this approach, one has to write an ansatz for the holographic correlator that is a sum of Witten diagrams. To do so, one has to consider the most general selection rules that follow from the structure of the underlying supergravity theory, while having some arbitrary coefficients in the ansatz. These coefficients are then fixed by imposing general consistency conditions on the correlator. 

Taking all of the above into consideration, in this work we take on ourselves to employ a position-space bootstrap for the computation of the $4$-point correlation functions of $\tfrac{1}{2}$-BPS operators of hypermultiplets in $\text{AdS}_2 \times \text{S}^2$ for arbitrary external charges. This task is, in some sense, a way to prove the emergence of a hidden $4$-dimensional conformal symmetry in this simple setup. To do so, we begin by writing down a general ansatz in terms of contact Witten diagrams with 0- and 2- derivatives corresponding to tree-level supergravity that is consistent with the general selection rules of the $\text{AdS}_2 \times \text{S}^2$ description. We, then, proceed to impose crossing symmetry, superconformal Ward identities and the bulk-point limit on our ansatz to determine the free coefficients. This fully fixes the answer up to an overall number. Our result agrees with the expectations from hidden conformal symmetry as we explicitly demonstrate.

The structure of this work is as follows: in \cref{sec: prems} we briefly review some basic facts about $\text{AdS}_2 \times \text{S}^2$ supergravity, the kinematics of $4$-point correlation functions in $1$-dimensional CFTs, fermionic Witten diagrams and the flat-space limit in the position-space approach, namely the bulk-point limit. Subsequently in \cref{sec: 1111} we demonstrate our algorithm in great detail for the lowest-lying holographic correlator, the $\langle \mathcal{O}_1 \mathcal{O}_1 \mathcal{O}_1 \mathcal{O}_1 \rangle$. We proceed to the discussion of more general charges in \cref{sec: pqrs}. \Cref{sec: conf_symm} contains a review of some basic statements about hidden conformal symmetry and we explicitly show the agreement of our results with expectations of this hidden structure. We conclude and offer some suggestions for future research in \cref{sec: epilogue}. In \cref{app: d.functions,app: dbar.functions} we provide the characteristic relations governing $D$- and $\bar{D}$-functions and the explicit form of $\bar{D}$-functions used in the $\langle \mathcal{O}_1 \mathcal{O}_1 \mathcal{O}_1 \mathcal{O}_1 \rangle$ case, for the reader's convenience, respectively. 
\section{Preliminaries}\label{sec: prems}
\subsection{\texorpdfstring{$\text{AdS}_2 \times \text{S}^2$}{ads2s2} supergravity in a nutshell}\label{sec: ads2overview}
In this work we are studying correlation functions in a $1$-dimensional theory dual to scattering in the $\text{AdS}_2 \times \text{S}^2$ background. The supergravity Kaluza-Klein spectrum in $\text{AdS}_2 \times \text{S}^2$ has been obtained in \cite{Michelson:1999kn,Lee:1999yu,Corley:1999uz}, see also \cite{Keeler:2014bra} for more recent related work. 

More specifically, this background can be derived from $11$-dimensional supergravity starting from $\text{AdS}_2 \times \text{S}^2 \times \text{T}^7$ and reducing the theory on $\text{T}^7$, while considering only the zero-modes on the torus. In terms of the bulk description, this approximation holds true when the radius of the torus is parametrically smaller than the radii of the $\text{AdS}_2$ and the $\text{S}^2$ with the latter two being equal in this instance. After the $\text{T}^7$ a further compactification on $\text{AdS}_2 \times \text{S}^2$ yields $4$-dimensional, $\mathcal{N}=2$ supergravity. Upon reduction on $\text{S}^2$ one obtains an infinite tower Kaluza-Klein states that are organised into representations of the $\mathfrak{su}(1,1|2)$ superalgebra.

For illustrative purposes, we present the brane-scan of the $\text{AdS}_2 \times \text{S}^2 \times \text{T}^7$ theory in \cref{table: ads2s2t7}: 
\begin{table*}[ht!]
\begin{center}
\begin{tabular}{ |c|c|c|c|c|c|c|c|c|c|c|c|c|}
 \hline
 &&&&&&&&&&\\[-0.95em] 
  								&$x^0$ 	& $x^1$ 	& $x^2$ 	& $x^3$ 	& $x^4$ 	& $x^5$ 	& $x^6$ 	& $x^7$ 	& $x^8$ 	& $x^9$ 		& $x^{10}$			\\ 
 \hline
 M$2$-brane 					& --- 	& $\bullet$ & $\bullet$ & $\bullet$ & --- 		& --- 		& $\bullet$ & $\bullet$ & $\bullet$ & $\bullet$		& $\bullet$			\\ 
 \hline
 M$2$-brane 					& --- 	& $\bullet$ & $\bullet$ & $\bullet$ & $\bullet$ & $\bullet$ & --- 		& --- 		& $\bullet$ & $\bullet$		& $\bullet$		\\
 \hline
 M$5$-brane 					& --- 	& $\bullet$ & $\bullet$ & $\bullet$ & $\bullet$ & --- 		& $\bullet$ & --- 		& --- 		& ---			& ---	\\
 \hline
 M$5$-brane 					& --- 	& $\bullet$ & $\bullet$ & $\bullet$ & ---		& $\bullet$ & --- 		& $\bullet$ & --- 		& ---			& ---			\\
 \hline
\end{tabular}
\caption{The supersymmetric brane intersection. In the above notation --- denotes that a brane extends along that particular direction, while $\bullet$ means that the coordinate is transverse to the brane.}
\label{table: ads2s2t7}
\end{center}
\end{table*}

It is worthwhile pointing out that the brane configuration presented in \cref{table: ads2s2t7} was originally discovered in \cite{Klebanov:1996mh} as a connection to $4$-dimensional black holes. The authors in \cite{Klebanov:1996mh} considered the dimensional reduction of the $11$-dimensional supergravity background in type IIA and subsequently performed a T-duality transformation to derive a type IIB supergravity background with differently arranged stacks of D$3$-branes. 

We proceed to describe some basic facts about the spectrum of the theory. We will mainly follow \cite{Keeler:2014bra}. 

The matter content of $4$-dimensional, $\mathcal{N}=2$ supergravity contains $1$ graviton, $6$ gravitinos, $15$ vector and $10$ (complex) hypermultiplets. The fields in $\text{AdS}_2 \times \text{S}^2$ are organised in terms of two quantum numbers, $h$ and $j$, with the former being the lowest eigenvalue of the generator of the $\text{SL}(2,\mathbb{R})$ and the latter the relevant number for the $\text{SU}(2)$. Hence, an $(h,j)$-representation has a $(2j+1)$-degeneracy from the $\text{SU}(2)$ and an infinite tower of states with eigenvalues $h, h+1, h+2, \ldots$ from the $\text{SL}(2,\mathbb{R})$. All fields are organised in chiral multiplets that assume the form: 
\begin{equation}\label{eq: chiral_multiplets}
(k,k), 
\quad
2\left(k+\tfrac{1}{2},k-\tfrac{1}{2}\right), 
\quad
(k+1,k-1)
\,		,
\end{equation}
with $k$ in the above taking values $k=\tfrac{1}{2},1,\tfrac{3}{2},2,\ldots$. Note that the case $k=\tfrac{1}{2}$ is special and the final term of \cref{eq: chiral_multiplets} should be understood as the empty representation. The chiral multiplets in \cref{eq: chiral_multiplets} are short multiplets. As it turns out, there is a unique way to organise the matter content of the theory into chiral multiplets as described by \cref{eq: chiral_multiplets}:
\begin{equation}
\begin{aligned}
\text{Supergravity multiplet:}
\qquad
&2 \left[(k+2,k-2), 2 \left(k+\tfrac{5}{2},k+\tfrac{3}{2}\right), (k+3,k+1) \right]
\,			,
\\
\text{Gravitino multiplet:}
\qquad
&2 \left[\left(k+\tfrac{3}{2},k+\tfrac{3}{2}\right), 2 (k+2,k+1), \left(k+\tfrac{5}{2},k+\tfrac{1}{2}\right) \right]
\,			,
\\
\text{Hypermultiplet:}
\qquad
&2 \left[\left(k+\tfrac{1}{2},k+\tfrac{1}{2}\right), 2 (k+1,k), \left(k+\tfrac{3}{2},k-\tfrac{1}{2}\right) \right]
\,			,
\\
\text{Vector multiplet:}
\qquad
&2 \left[(k+1,k+1), 2 \left(k+\tfrac{3}{2},k+\tfrac{1}{2}\right), (k+2,k) \right]
\,			,
\end{aligned}
\end{equation}
where $k=0,1,2,\ldots$ in the above. 
\subsection{Kinematics of four-point functions}\label{sec: kinematics}
We are interested in the $4$-point functions of $\tfrac{1}{2}$-BPS operators in a $1$-dimensional CFT. We will briefly review the formalism here following the discussion in \cite{Abl:2021mxo,Heslop:2023gzr}.

We start by noting that it is not possible to define a stress-energy tensor in a $1$-dimensional theory, since that would be just a constant. Therefore, we are examining bulk theories with no gravitational degrees of freedom. In these theories, however, we can construct correlation functions as they can be thought of as arising purely from the symmetries of the bulk picture and hence we can formally consider them as correlators of a CFT on the boundary of the space.

These $4$-point correlation functions admit a large central charge expansion, $c$. In the large-$c$ limit, the $\tfrac{1}{2}$-BPS operators are dual to scalars in the bulk $\text{AdS}_2$ that follow from the infinite KK tower of modes on the $\text{S}^2$. We, furthermore, want to address the low-energy limit of the theory. In this limit, the theory is $4$-dimensional, $\mathcal{N}=2$ supergravity. However, contrary to \cite{Abl:2021mxo}, we will not deal with sub-leading contributions that come as higher-derivative corrections. To account for this, we introduce a small parameter, $\alpha$, for which the $\alpha \rightarrow 0$ limit is the strict low-energy limit. 

Taking the above into consideration, we can define the double expansion to be given by $\alpha^{k-1} c^{-m}$, with $2k$ derivatives in the bulk scalar interaction and $k$ and $m$ being non-negative integers. More explicitly the expansion is
\begin{equation}
\underbrace{\alpha^0 c^0}_{\substack{\texttt{Generalised} \\ \texttt{free theory}}} 
+
\underbrace{\left(\alpha^0 c^{-1} + \overbrace{\alpha c^{-1} + \alpha^{2} c^{-1} + \ldots}^{\texttt{stringy corrections}} \right)}_{\text{AdS}~\texttt{tree-level}}   
+
\underbrace{\left(\alpha^0 c^{-2} + \overbrace{\alpha c^{-2} + \alpha^{2} c^{-2} + \ldots}^{\texttt{stringy corrections}} \right)}_{\text{AdS}~\texttt{$1$-loop}}  
+ 
\ldots
\end{equation}

The chiral primary fields have protected conformal dimension $\Delta$ and $\text{SU}(2)$ representation of spin-$j$ given by $j=\Delta$ with $\Delta=\frac{1}{2}, \frac{3}{2}, \frac{5}{2}, \ldots$. To keep a track of the R-symmetry structures it is useful to introduce $2$-component polarisation spinors $v^I$ such that the chiral primary fields are:
\begin{equation}\label{eq: chiral_fields_fermions}
 \psi_{I_1 \ldots I_{2\Delta} }v^{I_1} \ldots v^{I_{2\Delta}} = \psi_{\Delta} 
\,			,
\qquad
\qquad
\text{with}
\quad
\Delta=\frac{1}{2}, \frac{3}{2}, \frac{5}{2}, \ldots
\,			,
\end{equation}
where we can set the first component to $1$ such that $v^I=(1,y)$.

We are interested in the $4$-point correlation function of $\tfrac{1}{2}$-BPS operators that are described by \cref{eq: chiral_fields_fermions}. This $4$-point function is a correlator of fermionic primary fields, $\psi_{\Delta}$, with half-integer conformal dimensions and R-symmetry representations. It is convenient and useful to exchange this fermionic label for a bosonic one by considering the shift, $\Delta \rightarrow p-\tfrac{1}{2}$, with $p=1,2,\ldots$. Having done so, we label the primaries as $\mathcal{O}_{p}$. In this notation the $\tfrac{1}{2}$-BPS operators have dimensions and R-symmetry representations given by $p-\tfrac{1}{2}$ and the correlator is written as: 
\begin{equation}\label{eq: 4pt_fun_O_01}
\langle \mathcal{O}_{p_1}(x_1,y_1) \mathcal{O}_{p_2}(x_2,y_2) \mathcal{O}_{p_3}(x_3,y_3) \mathcal{O}_{p_4}(x_4,y_4) \rangle =   G_{p_i}(x_i,y_i) 
\,			.
\end{equation}

The $4$-point correlators of chiral primaries of the theory, \cref{eq: 4pt_fun_O_01}, can be written as functions of the conformal and R-symmetry cross-ratios. In a $1$-dimensional CFT there is only one conformal cross-ratio given by:
\begin{equation}\label{eq: def_of_x}
x = \frac{x_{12}x_{34}}{x_{13}x_{24}}
\,			, 
\end{equation}
that is related to the conformal cross-ratios in higher-dimensional theories via:
\begin{equation}\label{eq: cross-ratios}
U = z \bar{z} = \frac{x^2_{12}x^2_{34}}{x^2_{13}x^2_{24}}
=x^2
\,		,
\qquad
V = (1-z)(1-\bar{z}) = \frac{x^2_{23}x^2_{14}}{x^2_{13}x^2_{24}}=
(1-x)^2
\,		,
\end{equation}
and by setting
\begin{equation}
z = \bar{z} = x
\,			,
\end{equation}
we can see that the conformal cross-ratio of the $1$-dimensional theory corresponds to the holomorphic limit of the usual cross-ratios from higher-dimensions. We have used the abbreviation $\mathfrak{x}_{ab}$ for various quantities, which is defined as $\mathfrak{x}_{ab} = \mathfrak{x}_a - \mathfrak{x}_b$, unless otherwise stated. Similarly for $\text{SU}(2)$ R-symmetry we can define a cross-ratio $y$:
\begin{equation}\label{eq: def_of_y}
y = \frac{y_{12}y_{34}}{y_{13}y_{24}}
\,			.
\end{equation}
We note that the $x$ and $y$ can be understood as the bosonic components of the super-Grassmannian $\text{Gr}(1|1,2|2)$ matrix of coordinates \cite{Doobary:2015gia} that is relevant to the description of the correlation functions in analytic superspace. 

Having exchanged fermionic labels in favour of bosonic ones and introduced the appropriate cross-ratios above given by \cref{eq: def_of_x,eq: def_of_y}, we are able to re-write the $4$-point correlation function in \cref{eq: 4pt_fun_O_01} in the following manner:
\begin{equation}\label{eq: 4pt_fun_O}
G_{p_i}(x_i,y_i)  = \mathcal{P}_{\{p_i\}}~\mathcal{G}_{\{p_i\}}(x,y)
\,			,
\end{equation}
with 
\begin{equation}\label{eq: P_with_p}
\mathcal{P}_{\{p_i\}} = g^{p_1 + p_2 - 1}_{12} g^{p_3 + p_4 - 1}_{34} \left(\frac{g_{14}}{g_{24}}\right)^{p_{12}} \left(\frac{g_{13}}{g_{14}}\right)^{p_{34}}
\,	,
\end{equation}
and 
\begin{equation}
g_{ab} = \frac{y_{ab}}{x_{ab}}
\,			,
\end{equation}
and we remind the reader that $p_{ij}=p_i-p_j$ in \cref{eq: P_with_p}. 

The fermionic charges of the $\mathfrak{su}(1,1|2)$ impose more constraints, which are the superconformal Ward identities. In this notation they assume the form \cite{Abl:2021mxo}:
\begin{equation}\label{eq: ward_identities}
\frac{\partial}{\partial x} \mathcal{G}(x,y) \bigg|_{x=y} = 0 
\,				.
\end{equation}
The solution to the superconformal Ward identities, \cref{eq: ward_identities}, yields: 
\begin{equation}\label{eq: def_curly_defs}
\mathcal{G}_{\{p_i\}}(x,y) = \mathcal{G}_{0,\{p_i\}} + \mathcal{R} ~ \mathcal{H}_{\{p_i\}} 
\,          ,
\end{equation}
where in the above $\mathcal{G}_{0,\{p_i\}}$ denotes the protected piece, $\mathcal{R}$ is determined by superconformal symmetry to be:
\begin{equation}\label{eq: def_curly_R}
\mathcal{R} = 1 - \frac{x}{y}
\,          ,
\end{equation}
and $\mathcal{H}_{\{p_i\}}$ is the reduced correlator that carries the non-trivial dynamical information. 

We find it useful to re-write the above solution with all the kinematic factors being restored as: 
\begin{equation}\label{eq: straight_defs}
G_{\{p_i\}} =G_{0,\{p_i\}} + R ~ H_{\{p_i\}} 
\,          .
\end{equation}
The $R$ in \cref{eq: straight_defs} is related to $\mathcal{R}$ given by \cref{eq: def_curly_R} in the following way:
\begin{equation}
R = x_{13} x_{24} y_{12} y_{34} \mathcal{R}
\,          ,
\end{equation}
and from the above we can see that $R$ is crossing anti-symmetric reflecting properly the fermionic statistics. Finally, we can, also, work out the relation between the interacting parts with and without the kinematic factors. It reads: 
\begin{equation}
H_{\{p_i\}} = \frac{1}{x_{13}x_{24} y_{12} y_{34}} \mathcal{P}_{\{p_i\}} \mathcal{H}_{\{p_i\}}
\,          .
\end{equation}

Before closing this section and to set up concrete conventions, we mention that in this work, we will be assuming that the charges are in ascending order, namely $p_1 \leq p_2 \leq p_3 \leq p_4$, without loss of generality and we distinguish between two cases:
\begin{equation}\label{eq: cases_def}
p_1 + p_4 \geq p_2 + p_3
\,
\quad
\text{Case I}
\,
\qquad
\text{and}
\qquad
p_1 + p_4 < p_2 + p_3
\,
\quad
\text{Case II}
\,      .
\end{equation}
Correlators are characterised by their extremality, which we denote by $\mathcal{E}$, and for the two cases that we have distinguished above is given by:
\begin{equation}\label{eq: extremality_def}
\mathcal{E} = \frac{1}{2}(p_1+p_2+p_3-p_4)+1
\,
\quad
\text{for Case I}
\,          ,
\qquad
\text{and}
\qquad
\mathcal{E} = p_1 + 1 
\,
\quad
\text{for Case II}
\,          .
\end{equation}
These definitions will become useful at a later stage when discussing the bulk-point limit in \cref{sec: bulkpoint}.
\subsection{Fermionic Witten diagrams}\label{sec: fermiwitten}
For the purposes of our analysis, a pivotal role is played by contact Witten diagrams. These are depicted in \cref{fig: contact_witten}:
\begin{figure}[!ht]  
\centering 
\begin{subfigure}[b]{0.45\linewidth}
\begin{tikzpicture}
\coordinate (center) at (0,0);
\node[line width=0.5mm, node] (scalarblob) {};
\draw[>=latex,<-, very thick, blue] (center)
to 
node[right,label={[font=\large,text=black,label distance=0.9cm]80:$\psi_4$}] {} (scalarblob.north east);
\draw[>=latex,->, very thick, blue] (center)
to 
node[right,label={[font=\large,text=black,label distance=0.9cm]-80:$\bar{\psi}_3$}] {} (scalarblob.south east);
\draw[>=latex,<-, very thick, red] (center)
to 
node[right,label={[font=\large,text=black,label distance=-2.35cm]-57:$\bar{\psi}_1$}] {} (scalarblob.north west);
\draw[>=latex,->, very thick, red] (center)
to 
node[right,label={[font=\large,text=black,label distance=-2.35cm]57:$\psi_2$}] {} (scalarblob.south west);
\end{tikzpicture}
\caption{The $s$-channel fermionic contact Witten diagram.} \label{fig: schannel}  
\end{subfigure}
\hfill
\begin{subfigure}[b]{0.45\linewidth}
\begin{tikzpicture}
\coordinate (center) at (0,0);
\node[line width=0.5mm, node] (scalarblob) {};
\draw[>=latex,->, very thick, blue] (center)
to 
node[right,label={[font=\large,text=black,label distance=0.9cm]80:$\psi_4$}] {} (scalarblob.north east);
\draw[>=latex,<-, very thick, blue] (center)
to 
node[right,label={[font=\large,text=black,label distance=-2.35cm]-57:$\bar{\psi}_1$}] {} (scalarblob.north west);
\draw[>=latex,<-, very thick, red] (center)
to 
node[right,label={[font=\large,text=black,label distance=-2.35cm]57:$\psi_2$}] {} (scalarblob.south west);
\draw[>=latex,->, very thick, red] (center)
to 
node[right,label={[font=\large,text=black,label distance=0.9cm]-80:$\bar{\psi}_3$}] {} (scalarblob.south east);
\end{tikzpicture}
\caption{The $u$-channel fermionic contact Witten diagram.} \label{fig: uchannel}  
\end{subfigure}
\hfill
\begin{subfigure}[b]{0.45\linewidth}
\begin{tikzpicture}
\coordinate (center) at (0,0);
\node[line width=0.5mm, node] (scalarblob) {};
\draw[-, very thick, orange] (center)
to 
node[right,label={[font=\large,text=black,label distance=0.9cm]80:$\phi_4$}] {} (scalarblob.north east);
\draw[-, very thick, orange] (center)
to 
node[right,label={[font=\large,text=black,label distance=0.9cm]-80:$\phi_3$}] {} (scalarblob.south east);
\draw[-, very thick, orange] (center)
to 
node[right,label={[font=\large,text=black,label distance=-2.35cm]-53:$\phi_1$}] {} (scalarblob.north west);
\draw[-, very thick, orange] (center)
to 
node[right,label={[font=\large,text=black,label distance=-2.35cm]57:$\phi_2$}] {} (scalarblob.south west);
\end{tikzpicture}
\caption{Tree-level contact Witten diagram of external scalars.} \label{fig: scalar_diagram}  
\end{subfigure}
\caption{Tree-level contact Witten diagrams. In \cref{fig: schannel} we depict the $s$-channel diagram of external fermions, and in \cref{fig: uchannel} the relevant $u$-channel diagram. In \cref{fig: scalar_diagram} we draw a scalar contact Witten diagram.  Note that in the case of fermions, unlike the associated scalar Witten diagrams, there is no $t$-channel contribution.} 
\label{fig: contact_witten}  
\end{figure}
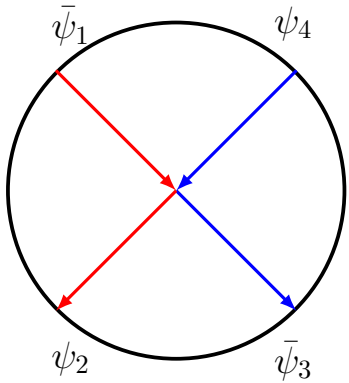
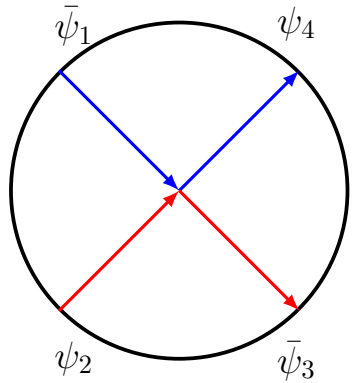
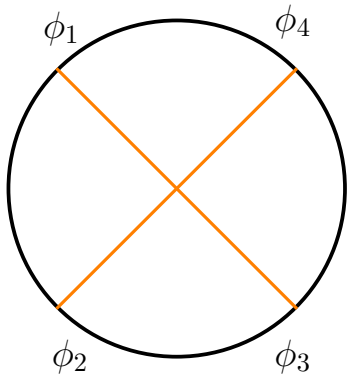 

The tree-level contact Witten diagam of external scalars, shown in \cref{fig: scalar_diagram}, that carry dimensions $\Delta_1, \ldots, \Delta_4$ and with no derivatives is represented in terms of the so-called $D$-function, which is given by: 
\begin{equation}
D_{\Delta_1 \Delta_2 \Delta_3 \Delta_4} = \int d^2 X  K_{\Delta_1}(X,P_1) K_{\Delta_2}(X,P_2) K_{\Delta_3}(X,P_3) K_{\Delta_4}(X,P_4)
\,          ,
\end{equation}
with the scalar bulk-to-boundary propagator being equal to: 
\begin{equation}
K_{\Delta}(X,P)=\frac{1}{(- 2 X \cdot P)^{\Delta}}
\,          .
\end{equation}

In \cite{Carmi:2021dsn} the author studied various classes of Witten diagrams that contain different number of external fermions, see for instance \cref{fig: schannel,fig: uchannel}. We are interested in the case that schematically is written as $\langle\bar{\psi} \psi \bar{\psi} \psi\rangle$. The main result related to our purposes here, is that these diagrams are essentially proportional to the associated scalar Witten diagrams. We briefly review some basic features leading to that conclusion and refer the interested reader to \cite[section 2.1]{Carmi:2021dsn} for a more thorough exposition\footnote{Note that in \cite{Carmi:2021dsn} the author has considered a more general case with external fermions and a scalar $2$-point bulk correlator exchanged. Following the notation of that paper, for our analysis, it is sufficient to set $S(x_1,x_2)$ to be a $\delta$-function.}. Before proceeding, however, we feel it necessary to make some remarks on the embedding of spinors; see \cite{Isono:2017grm,Iliesiu:2015qra} for details. 

Let us start by passing from the physical $1$-dimensional spacetime to the $3$-dimensional embedding space. Working in embedding space is very convenient. The action of the special conformal transformations is realised non-linearly by the coordinates, $x^{\mu}$, of the phyiscal spacetime. However, the embedding space coordinates $P^A$ transform linearly under the special conformal transformations. We consider $\mathcal{M}$ to be a $1$-dimensional Euclidean spacetime with metric $\eta_{\mu \nu}$ and we call the embedding space $\mathbb{M}$ endowed with the flat metric $\eta_{AB}=\text{diag}(1,1,-1)$. The embedding of $\mathcal{M}$ into $\mathbb{M}$ is realised as the null hypersurface $P^2 = \eta_{AB} P^A P^B =0$. We can introduce the light-cone coordinates as $P^{\pm} = P^2 \pm P^1$, such that we write the $3$-dimensional coordinates in embedding space as $P\equiv(P^{\mu},P^+,P^-)=(x^{\mu},1,x^2)$. 

Now, we wish to consider a spinor, $\psi(x)$ in the physical space that is a primary field. Formally speaking, the spinor representation is Majorana. We can take the $\gamma$-matrices to be real and hence the Majorana spinor has real components. For the $\psi(x)$ we can consider a spurionic field $s$ that is position-independent such that we form $\psi(x,s)=\bar{s}\psi(x)$, that is now a spacetime scalar\footnote{In a $1$-dimensional theory, the Grassmann-even spinor is a constant that we can rescale to $1$. We wrote the expression here formally for clarity retaining $s$ explicit, and in the remaining of our work we will set it to $1$.}. Working in the same vein, we can form a spacetime scalar starting from a spinor field in the embedding space as $\Psi(P,S) = \bar{S} \Psi(P)$. The relation between the two spacetime scalars formed out of spinors in the physical and embedding spaces is:
\begin{equation}
\psi(x,s) = \Psi(P,S)
\,      ,
\end{equation}
while the relation of the polarisation spinors in the two spaces reads\footnote{As we mentioned earlier we have already set $s=1$ when expressing \cref{eq: S_barS}. For a more explicit relation that holds between a $d$-dimensional physical space and a $(d+2)$-dimensional embedding space see \cite{Isono:2017grm}.}: 
\begin{equation}\label{eq: S_barS}
S = \begin{pmatrix}
    x\\
    1
    \end{pmatrix}
\,      ,
\qquad
\bar{S} = (1, -x)
\,      .
\end{equation}
The $2$-point function of the spinors is given by \cite{Isono:2017grm,Iliesiu:2015qra}:  
\begin{equation}
\langle \Psi(P_1,S_1) \bar{\Psi}(P_2,S_2) \rangle =  \frac{\langle \bar{S}_1 S_2\rangle}{P^{2 \Delta_{\psi}+\tfrac{1}{2}}_{12}}
\,          .
\end{equation}
Using \cref{eq: S_barS} we can obtain 
\begin{equation}
\langle \psi(x_1) \bar{\psi}(x_2) \rangle = - \frac{x_{12}}{x^{2 \Delta_{\psi}+1}_{12}}
\end{equation}
as expected. 

Now, we turn our attention to the $4$-point function of $4$ external fermions with scaling dimensions $\Delta_i$ with $i=1,\ldots,4$:
\begin{equation}\label{eq: fermi_witten_4}
\Psi_{pqrs} \equiv  \langle\bar{\Psi}_{\Delta_1}(P_1,S_1) \Psi_{\Delta_2}(P_2,S_2) \bar{\Psi}_{\Delta_3}(P_3,S_3) \Psi_{\Delta_4}(P_4,S_4) \rangle
\,			,
\end{equation}
that arises from contact interactions in AdS. This fermionic $4$-point function can be constructed by 
the fermionic bulk-to-boundary propagators which are given by: 
\begin{equation}
\begin{aligned}
&\mathcal{K}^{(X)}_{a} \equiv \mathcal{K}_{\Delta}(X,\bar{S}_{a \mathbf{b}},P_a,S_{a \partial}) =  \frac{\langle \bar{S}_{\mathbf{b}} S_{\partial} \rangle}{(- 2 X \cdot P_a)^{\Delta_a+\tfrac{1}{2}}}
\,          ,
\\
&\bar{\mathcal{K}}^{(X)}_{a} \equiv \mathcal{K}_{\Delta}(X,S_{a \mathbf{b}},P_a,\bar{S}_{a \partial})
= 
 \frac{\langle \bar{S}_{\partial} S_{\mathbf{b}} \rangle}{(- 2 X \cdot P_a)^{\Delta_a+\tfrac{1}{2}}}
\,          ,
\end{aligned}
\end{equation}
where in the above $\mathbf{b}$ stands for bulk and $\partial$ denotes the boundary, and we have introduced the polarisation spinor in the bulk similarly to the discussion above. 
The answer of the $4$-point function contains $2$ pieces, the $s$- and $u$-channel parts; see \cref{fig: schannel,fig: uchannel}
\begin{equation}
\Psi_{pqrs} = F_{12|34} + F_{14|23}
\,          .
\end{equation}
For concreteness, we focus on the $s$-channel part, since the answer for the $u$-channel follows straightforwardly. We have \cite{Carmi:2021dsn}
\begin{equation}\label{eq: F_1234}
F_{12|34} = \int d^2 X 
(\partial_{S_{1b}} \partial_{\bar{S}_{2b}}) 
\bar{\mathcal{K}}^{(X)}_{1} \mathcal{K}^{(X)}_{2}
(\partial_{S_{3b}} \partial_{\bar{S}_{4b}}) 
\bar{\mathcal{K}}^{(X)}_{3} \mathcal{K}^{(X)}_{4}
\,          ,
\end{equation}
with the integral being over $\text{AdS}_2$. Using the identity \cite{Carmi:2018qzm,Carmi:2021dsn}\footnote{It is worthwhile mentioning that the relation of external fermions with a scalar exchange to external scalars with a scalar exchange with a shift by $\tfrac{1}{2}$ in the conformal dimensions had been noticed in \cite{Kawano:1999au,Faller:2017hyt} as well. The authors in \cite{Carmi:2018qzm} stressed that whether we have an exchange of an elementary or a composite scalar state does not make a difference in the arguments.}
\begin{equation}
(\partial_{S_{1b}} \partial_{\bar{S}_{2b}}) 
\bar{\mathcal{K}}^{(X)}_{1} \mathcal{K}^{(X)}_{2} 
=
 x_{12}
K_{\Delta_1 + \tfrac{1}{2}}(X,P_1) K_{\Delta_2 + \tfrac{1}{2}}(X,P_2)
\,          ,
\end{equation}
that relates the fermionic bulk-to-boundary propagators to their scalar counterparts we are able to express \cref{eq: F_1234} as: 
\begin{equation}\label{eq: F_1234_bosonic}
F_{12|34} =  x_{12} x_{34} \int d^2 X  K_{\delta_1}(X,P_1) K_{\delta_2}(X,P_2) K_{\delta_3}(X,P_3) K_{\delta_4}(X,P_4)
=
x_{12} x_{34}
D_{\delta_1 \delta_2 \delta_3 \delta_4}
\,          ,
\end{equation}
having defined $\delta_i = \Delta_i + \tfrac{1}{2}$. Hence, we observe that \cref{eq: F_1234} is re-written completely in terms of scalar propagators with shifted conformal dimensions; namely a $D$-function with shifted weights. 
\subsection{The bulk point limit}\label{sec: bulkpoint}
Another important ingredient that we are going to utilize in our bootstrap algorithm is the flat-space limit of an AdS amplitude as described by the relevant correlator in the dual CFT. In the position-space representation that we have employed in this work, it amounts to considering the so-called bulk-point limit\footnote{
The flat-space limit can, of course, be also formulated in the Mellin-space representation. This has been developed in \cite{Penedones:2010ue,Fitzpatrick:2011hu} and has proved itself to be a very useful resource that constrains correlators at tree-level including string corrections as well \cite{Goncalves:2014ffa,Alday:2018pdi,Binder:2019jwn,Drummond:2019odu,Drummond:2020dwr,Aprile:2020mus} and also at one-loop \cite{Alday:2019nin,Alday:2020tgi,Alday:2021ajh} and two-loop corrections \cite{Drummond:2022dxw,Huang:2021xws}. An $\text{AdS}_5 \times \text{S}^5$ tailored formalism has been proposed in \cite{Aprile:2020luw}.
}. The bulk-point limit in its essence is the statement that if one considers a sufficiently localised AdS wave-packet, one can focus on a point in the bulk. Effectively, by doing so one cannot see any effects of the curvature, and thus this recovers the scattering amplitude in flat-space. 
We will closely follow \cite{Drummond:2022dxw} in taking the bulk-point limit of AdS amplitudes. 

More precisely the bulk-point limit requires to analytically continue from Euclidean to Lorentzian signature, which in terms of cross-ratios amounts to considering an analytic continuation of $z$ around $0$ counter-clockwise and $\bar{z}$ around $1$, also counter-clockwise. After that, taking the limit $\bar{z} \rightarrow z$ gives a singularity of the schematic form:
\begin{equation}
\mathcal{H}(z,\bar{z}) 
\xrightarrow[z \rightarrow \bar{z}]{\substack{\texttt{analytic continuation} \\ \texttt{in} ~ z, \bar{z}}}
\frac{\mathcal{F}(z)}{(z-\bar{z})^p}
\,              .
\end{equation}
Such a behaviour is the expected one for any holographic correlation function possessing a local bulk-dual description \cite{Gary:2009ae,Heemskerk:2009pn,Okuda:2010ym,Maldacena:2015iua}. The residue of the singularity is related to the $4$-dimensional scattering amplitude of hypermultiplets in flat-space, $\mathcal{A}^{(4)}$ as:
\begin{equation}
\mathcal{A}^{(4)} \propto s^{\mathfrak{k}} \frac{\mathcal{F}(z)}{z^{\mathfrak{l}}(1-z)^{\mathfrak{m}}}
\,			,
\end{equation}
where $\mathfrak{k},\mathfrak{l},\mathfrak{m}$ are integers and are related to the dimension of the bulk interaction vertex; see for example \cite{Drummond:2022dxw,Heemskerk:2009pn}. The parameter $z$ is dimensionless and is defined in terms of the scattering angle $\theta$ or in terms of the Mandelstam parameters via: 
\begin{equation}
z \equiv \frac{s+t}{s} = \frac{1}{2}(1+\cos\theta)
\,			.
\end{equation}

We wish to pause for a moment, in order to make a clarifying comment. In a $1$-dimensional theory there is only one conformal cross-ratio, and hence considering the analytic continuation is a bit tricky. However, in this theory we can still use the higher-dimensional prescription described above. Our correlator is a sum made of $D$-functions, or equivalently $\bar{D}$-functions using \cref{eq: funcDbar_def}. The correct thing to do is to first perform the analytic continuations independently in terms of the higher-dimensional cross-ratios, prior to taking the limit $z \rightarrow \bar{z}$. 

Since we are dealing with a sum of $\bar{D}$-functions essentially, we need to understand what the individual contribution of a given $\bar{D}$-function is in the bulk-point limit\footnote{We are extremely grateful to Hynek Paul for related, clarifying discussions at this point.}. In order to take the $z \rightarrow \bar{z}$ limit, we recall that any $\bar{D}$-function can be uniquely decomposed as: 
\begin{equation}\label{eq: d_bar_decompose}
\bar{D}_{\Delta_1,\Delta_2,\Delta_3,\Delta_4} = \mathcal{R}_{\phi} \phi^{(1)}(z,\bar{z}) + \mathcal{R}_U \log U + \mathcal{R}_V \log V + \mathcal{R}_{1}
\,			,
\end{equation}
where in the above the various $\mathcal{R}$ denote rational functions of $z$ and $\bar{z}$ and $\phi^{(1)}(z,\bar{z})$ is the well-known scalar one-loop box integral in four dimensions which is evaluated in terms of dilogarithms, see \cref{eq: phi_1_def} for its precise form. Upon taking the bulk-point limit in \cref{eq: d_bar_decompose} the contributions of the $\{\log U,\log V,1\}$ are sub-leading compared to the $\phi^{(1)}(z,\bar{z})$ part. Hence, we can write the schematic, but suggestive:
\begin{equation}\label{eq: bulk_point_dbar}
\bar{D}_{\Delta_1,\Delta_2,\Delta_3,\Delta_4} \xrightarrow[\texttt{limit}]{\texttt{bulk-point}} \{\phi^{(1)}(z,\bar{z}),\cancel{\log U},\cancel{\log V},\cancel{1}\}
\,			,
\end{equation}

In backgrounds of the form $\text{AdS} \times \text{S}$ we, also, have to account properly for the higher KK excitations in the internal manifold. The discussion so far applies to the lowest-lying KK mode. For the higher KK states, the result of the flat-space limit is that of the $\text{AdS}$ times a factor that accounts for the KK modes and depends on the polarisations. 
\begin{equation}
\mathcal{A}^{(4)}_{\{p_i\}} \propto \mathcal{B}_{\{p_i\}} \mathcal{A}^{(4)}_{1111}
\end{equation}
To write down the $\mathcal{B}_{\{p_i\}}$ factor we need to regroup the spinors appearing in the definition of the chiral primary spinors fields, \cref{eq: chiral_fields_fermions}, in such a way that we form $\text{SO}(3)$ null vectors $t^{IJ}_i = v^I_i \cdot v^J_i$ that satisfy $t_i \cdot t_i =0$. The result is then given by the Wick contractions of the $p_i-1$ null vectors, $t_i$, of the $\text{SO}(3)$ and is proportional to \cite{Alday:2021odx}:
\begin{equation}\label{eq: polarisation_flat_space}
\mathcal{B}_{\{p_i\}} \propto \prod_{i<j} (t_{ij})^{-c_{ij}} ~ (t_{12}t_{34})^{2 - \mathcal{E}} ~ \texttt{Wick}\big[\underbrace{t_1\ldots t_1}_{p_1 - 1} \ldots \underbrace{t_4\ldots t_4}_{p_4 - 1}\big]
\,          ,
\end{equation}
where $t_{ij} = t_i \cdot t_j = (y_{ij})^2$ and with the exponents being given by: 
\begin{equation}
\begin{aligned}
c^0_{12} &= c^0_{13} = 0
\,          ,
\\
c^0_{34} &= \frac{1}{2}|p_3+p_4-p_1-p_2|
\,          ,
\\
c^0_{24} &= \frac{1}{2}|p_2+p_4-p_1-p_3|
\,          ,
\\
c^0_{14}    &= \frac{1}{2}|p_1 + p_4 - p_2 - p_3|
\,      ,
\quad
c^0_{23}    = 0
\,      ,
\quad
\text{for Case I}
\,      ,
\\
c^0_{14}    &= 0
\,      ,
\quad
c^0_{23}    = \frac{1}{2}|p_1 + p_4 - p_2 - p_3|
\,      ,
\quad
\text{for Case II}
\,      .
\end{aligned}   
\end{equation}

In terms of the R-symmetry cross-ratio \cref{eq: polarisation_flat_space} is given by\footnote{To facilitate the interested reader we note that in the language of \cite{Alday:2021odx} their internal cross-ratios $\sigma$ and $\tau$ can be realised in terms of the $y$ in our setup as $\sigma = \frac{1}{y^2}$ and $\tau = \left(\frac{1-y}{y}\right)^2$.}: 
\begin{equation}\label{eq: factor_B}
\mathcal{B}_{\{p_i\}} = \sum_{\substack{i+j+k=\mathcal{E}-2 \\ 0 \leq i,j,k \leq \mathcal{E}-2}} \frac{(1-y)^{2j}}{y^{2(i+j)}} \frac{1}{i!j!k! \left(i + \frac{|p_3 + p_4 - p_1 - p_2|}{2} \right)! \left(j + \frac{|p_1 + p_4 - p_2 - p_3|}{2} \right)! \left(k + \frac{|p_2 + p_4 - p_1 - p_3|}{2} \right)!}
\,          .
\end{equation}

We remind the readers that the meaning of Cases I and II is related to the ordering of the external charges and was spelled out in \cref{eq: cases_def}, while our definition of extremality for these two cases is given by \cref{eq: extremality_def}.
\section{The simplest bootstrap: the \texorpdfstring{$\langle \mathcal{O}_1 \mathcal{O}_1 \mathcal{O}_1 \mathcal{O}_1 \rangle$}{o1o1o1o1} correlator}\label{sec: 1111}
We begin by considering the AdS scattering of the lowest-lying states, the $\mathcal{O}_1$ operators, and making an ansatz in terms of contact Witten diagrams for the $\langle \mathcal{O}_1 \mathcal{O}_1 \mathcal{O}_1 \mathcal{O}_1 \rangle$ correlator. In the ansatz we allow for all structures that can appear with $0$- and $2$-derivatives. The ansatz is the following\footnote{We denote the ansatz in terms of the $x_i$ and $y_i$ by $A_{\{p_i\}}$ and the one written in terms of cross-ratios by $\mathcal{A}_{\{p_i\}}$ to have consistent notation with \cref{sec: kinematics}.}: 
\begin{equation}\label{eq: ansatz_1111}
    \begin{aligned}
A_{1111} &= \sum_{i} \left(\lambda_{1,i} y_{12} y_{34} + \lambda_{2,i} y_{13} y_{24}  +  \lambda_{3,i} y_{14} y_{23}  \right) \Upsilon_i
\,      ,
\\
\rm{with}
\,      
\\
\{\Upsilon_i\} = 
\{
&x_{12} x_{34} D_{1111}, ~ x_{13} x_{24} D_{1111}, ~ x_{14} x_{23} D_{1111}; 
\\
&x_{12}x_{34}^3 D_{1122}, ~ x_{13}x_{24} x^2_{34} D_{1122}, ~  x_{14}x_{23} x^2_{34} D_{1122}; 
\\
&x_{13}x_{24}^3 D_{1212}, ~ x_{13}x_{24}^2 x_{34} D_{1212}, ~  x_{14}x_{23} x^2_{24} D_{1212};
\\
&x_{23}x_{14}^3 D_{2112}, ~ x_{13}x_{14}^2 x_{24} D_{2112}, ~  x_{12}x_{34} x^2_{14} D_{2112}
\}
\,      .
    \end{aligned}
\end{equation}
In the above, \cref{eq: ansatz_1111}, the first line of $\Upsilon_i$ contains all the $0$-derivative terms, while the remaining ones are the $2$-derivative structures.

Before we proceed to bootstrap \cref{eq: ansatz_1111} we wish to explain the  R-symmetry structures, the factors of $y$, and how they arise. R-symmetry requires that the polarisation spinors can only appear as polynomials 
\begin{equation}\label{eq: r_sym_aux_01}
\prod_{i<j} \left(v_{ij}\right)^{a_{ij}} = \prod_{i<j} \left(y_{ij}\right)^{a_{ij}}
\,          ,
\end{equation}
with $i$ and $j$ being particle numbers, and $a_{ij}$ being symmetric, $a_{ij} = a_{ji}$. Additionally, all the diagonal elements are being given by $a_{ii}=0$. Further, the exponents are non-negative, $a_{ij} \geq 0$. Furthermore, the $a_{ij}$ need to satisfy
\begin{equation}\label{eq: r_sym_aux_02}
\sum_{j} a_{ij} = 2 p_i -1
\,          .
\end{equation}
The integer solutions to the above constraints give all the R-symmetry structures in \cref{eq: ansatz_1111} and all subsequent examples we consider in \cref{sec: pqrs}.

Having sufficiently discussed all the terms that enter the ansatz we wrote above for the $\langle \mathcal{O}_1 \mathcal{O}_1 \mathcal{O}_1 \mathcal{O}_1 \rangle$ correlator, see \cref{eq: ansatz_1111}, we start by counting how many free coefficients enter in the ansatz and we observe that it comes with $36$ unfixed parameters. 

We are now at a position to implement crossing symmetry. There are $6$ ways to cross the correlator, however, only $3$ of them are independent. In the ansatz written in terms of the $x_i$, $y_i$ and the various $D$-functions, we consider the conditions:
\begin{equation}\label{eq: crossing_01}
A_{1111} + A^{(1 \leftrightarrow 2)}_{1111} = 0
\,          ,
\quad
A_{1111} + A^{(1 \leftrightarrow 3)}_{1111} = 0
\,          ,
\quad
A_{1111} + A^{(1 \leftrightarrow 4)}_{1111} = 0
\,          ,
\end{equation}
where in the above $A^{(1 \leftrightarrow 2)}_{1111}$ is the ansatz when considering crossing $1 \leftrightarrow 2$ and likewise for the rest. Note, also, that the ansatz is minus itself after crossing reflecting the fermionic statistics.

To implement the crossing conditions, one extracts a kinematic factor to re-write it as a functions of cross-ratios $x$ and $y$:
\begin{equation}\label{eq: aux_1111}
A_{1111} = \frac{y_{12}y_{34}}{x_{12}x_{34}} \mathcal{A}_{1111}
\,          .
\end{equation}  
Furthermore, one needs to use the explicit expressions for the $\bar{D}$-functions. Note that in higher dimensions any $\bar{D}$-function can be uniquely decomposed in the basis of \\ $\{\phi^{(1)}(z,\bar{z}),\log U,\log V, 1\}$. In the $1$-dimensional case one further needs to take the limit $z = \bar{z}=x$, and the basis reduces to $\{\log x, \log (1-x) ,1 \}$. The crossing conditions should hold for the coefficients of each element of the basis and for any values of $y$. The solution to the crossing equations, provides us with $10$ conditions on the free coefficients that we had in the ansatz. Note that these $10$ conditions relate the various free parameters amongst themselves.

Having obtained the solutions to the crossing symmetry equations, we wish to examine the implications of the superconformal Ward identities on the correlator. Having already the ansatz in terms of cross-ratios \cref{eq: aux_1111}, we can work as we did for crossing symmetry and write the ansatz in the basis spanned by $\{\log x, \log (1-x) ,1 \}$, and we wish to impose the superconformal Ward identities given by \cref{eq: ward_identities}. Note that the Ward identities should hold for the coefficients of each element of the basis independently. This gives us one and final condition on the undetermined parameters. 

After imposing this last condition on top of the previous ones coming from crossing symmetry on our ansatz, we obtain the answer: 
\begin{equation}
\mathcal{G}_{1111}  = \pi x \left(1 - \frac{x}{y} \right) ~ \bar{D}_{1111}
\,			,
\end{equation}
and hence we have fully fixed the correlator up to an overall number. This result agrees with the one derived in \cite{Abl:2021mxo}. Note that the protected part of the correlator $\mathcal{G}_{0,1111}=0$ which is a non-trivial statement.
\section{More examples with higher charges}\label{sec: pqrs}
As we saw in the previous section, the combined power of superconformal symmetry and crossing symmetry is sufficient to fix the lowest-lying $4$-point function up to an overall number. When increasing the external weights of the correlator, this is no longer true and we have to use, in addition to the above, the constraints that are provided by the bulk-point limit.  In order to demonstrate how this works, we exemplify the logic using the $\langle \mathcal{O}_2 \mathcal{O}_2 \mathcal{O}_2 \mathcal{O}_2 \rangle$ correlator. 

As we did in the previous example, \cref{sec: 1111}, we start by making an ansatz given as the linear combination of all possible $D$-functions that can appear with $0$- and $2$-derivatives:
\begin{equation}\label{eq: ansatz_2222}
    \begin{aligned}
A_{2222} &= \sum_{i} \left(\lambda_{1,i} y^3_{14} y^3_{23} y^2_{34} + \lambda_{2,i} y_{13} y^2_{14} y^2_{23} y_{24}  +  \lambda_{3,i} y_{13}^2 y_{14} y_{23} y^2_{24} + \ldots \right) \Upsilon_i
\,      ,
\\
\rm{with}
\,      
\\
\{\Upsilon_i\} &= \{x_{12} x_{34} D_{2222}, ~ x_{13} x_{24} D_{2222}, ~ x_{14} x_{23} D_{2222}; ~ x_{12}x_{34}^3 D_{2233}, ~ x_{13}x_{24} x^2_{34} D_{2233}, ~ \ldots  \}
\,      ,
    \end{aligned}
\end{equation}
and we remind the reader that all R-symmetry structures that enter in the above can be derived by combining \cref{eq: r_sym_aux_01,eq: r_sym_aux_02}.

In the ansatz we wrote above, \cref{eq: ansatz_2222}, we have $120$ unfixed parameters. The plan is to proceed in a similar manner as we did in \cref{sec: 1111}, namely use the power of crossing symmetry and then the implications of superconformal Ward identities. Subsequently, we will use the constraints of the bulk-point limit. Taking the bulk-point limit gives the last condition that we need, and upon imposing all the conditions we have 

We have sufficiently described earlier how to properly implement crossing symmetry and the Ward identities, and hence in this section we will proceed much faster. We point out that after solving the $3$ crossing symmetry equations, namely $1 \leftrightarrow 2$, $1 \leftrightarrow 3$ and $1 \leftrightarrow 4$ we obtain $20$ conditions on the free coefficients that we had in the ansatz. Furthermore, this time from the superconformal Ward identities, we obtain $2$ conditions on the free coefficients of the ansatz. Note that unlike the $\langle \mathcal{O}_1 \mathcal{O}_1 \mathcal{O}_1 \mathcal{O}_1 \rangle$ case, here the above conditions are not sufficient to fully fix the correlator. Particularly, the ansatz still contains $2$-derivative $D$-functions.

In order to fix the correlator in this case, we turn our attention to the bulk-point limit. The implementation of the bulk-point limit has been sufficiently discussed \cref{sec: bulkpoint}. Here just implement the steps described in that section for the various $\bar{D}$-functions that appear in the ansatz. The bulk-point limit gives the last condition on the free coefficients of the ansatz. The bulk-point limit has two important consequences on the correlator. The first one is that it fixes all the coefficients in front of the $2$-derivative $\bar{D}$-functions such that they cancel. The second one is that it uniquely determines the R-symmetry structures to be given by the $\mathcal{B}_{\{p_i\}}$-factor, given by \cref{eq: factor_B}. The final result is:
\begin{equation}
\mathcal{G}_{2222} =  \pi x^3 \left(1 - \frac{x}{y} \right)  2 \left(1 - \frac{1}{y} + \frac{1}{y^2} \right) ~ \bar{D}_{2222}
\,          .
\end{equation}
We note that the protected part of the correlator $\mathcal{G}_{0,2222}=0$ as was the case for the $\langle \mathcal{O}_1 \mathcal{O}_1 \mathcal{O}_1 \mathcal{O}_1 \rangle$.

We have checked explicitly that our position-space algorithm agrees with the all examples that are listed in \cite[equation.(74)]{Abl:2021mxo}. We have explicitly bootstrapped, in addition to these, the correlators $\langle \mathcal{O}_1 \mathcal{O}_1 \mathcal{O}_4 \mathcal{O}_4 \rangle$, $\langle \mathcal{O}_1 \mathcal{O}_1 \mathcal{O}_5 \mathcal{O}_5 \rangle$, $\langle \mathcal{O}_2 \mathcal{O}_2 \mathcal{O}_4 \mathcal{O}_4 \rangle$, $\langle \mathcal{O}_4 \mathcal{O}_4 \mathcal{O}_4 \mathcal{O}_4 \rangle$, and $\langle \mathcal{O}_3 \mathcal{O}_3 \mathcal{O}_3 \mathcal{O}_5 \rangle$. In all these examples we find the same structure. In particular the protected part of the correlator vanishes and the final result can be conveniently written as:
\begin{equation}\label{eq: bootstrap_general_charges}
\begin{aligned}
\mathcal{H}_{1122}
&= \pi x \bar{D}_{1122} 
\,      ,
\\
\mathcal{H}_{1133}
&= \pi x \frac{1}{2} \bar{D}_{1133}
\,      ,
\\
\mathcal{H}_{1144}
&= \pi x \frac{1}{6} \bar{D}_{1144}
\,      ,
\\
\mathcal{H}_{1155}
&= \pi x \frac{1}{24} \bar{D}_{1155}
\,      ,
\\
\mathcal{H}_{2233}
&= \pi x^3 \frac{3}{2} \left(1 - \frac{4}{3y} + \frac{4}{3y^2} \right) \bar{D}_{2233}
\,      ,
\\
\mathcal{H}_{2244}
&= \pi x^3 \frac{2}{3} \left(1 - \frac{2}{2y} + \frac{3}{2y^2} \right) \bar{D}_{2244}
\,      ,
\\
\mathcal{H}_{3333}
&= \pi x^5 \frac{3}{2} \left(1-\frac{2}{y}+\frac{3}{y^2}-\frac{2}{y^3}+\frac{1}{y^4}\right) \bar{D}_{3333} 
\,      ,
\\
\mathcal{H}_{4444}
&= \pi x^7 \frac{5}{9} \left(1 - \frac{3}{y} + \frac{33}{5y^2} - \frac{41}{5y^3} + \frac{33}{5y^4} - \frac{3}{y^5} + \frac{1}{y^6} \right) \bar{D}_{4444}
\,      ,
\\
\mathcal{H}_{3335}
&= \pi x \left(1 - \frac{1}{y} + \frac{1}{y^2} \right) \bar{D}_{3335}
\,      .
\end{aligned}
\end{equation}
As we will see, this structure is an implication of the underlying hidden conformal symmetry. 
\section{Hidden conformal symmetry}\label{sec: conf_symm}
In \cite{Rastelli:2016nze,Rastelli:2017udc} the authors obtained the $4$-point function of $\tfrac{1}{2}$-BPS operators with arbitrary external weights in the $4d$, $\mathcal{N}=4$ super Yang-Mills theory in the limit $N \rightarrow \infty$ and $\lambda = g^2 N \gg 1$. This result was obtained by solving an algebraic bootstrap problem. The remarkable simplicity of the formula hinted for some underlying principle governing this structure. In addition to this, there was further suggestive evidence for a hidden conformal symmetry based on the work of \cite{Aprile:2018efk}. This work studied the matrix of anomalous dimensions describing the mixing of double-trace operators constructed from different harmonics in the $\text{S}^5$. The eigenvalues of the problem are simple rational numbers for which a general formula was obtained.

The status of hidden conformal symmetry was further elaborated and made precise in \cite{Caron-Huot:2018kta}. The authors conjectured the existence of a $10$-dimensional conformal symmetry, in terms of which the $4$-point function of all $\tfrac{1}{2}$-BPS operators can be organized into one generating function. The latter is obtained by promoting the distances in $4$ dimensions to $10$-dimensional distances in the lowest-weight correlator. While we should mention that to this day we still lack a very rigorous explanation pertaining to the origin of such a symmetry, several intuitive arguments we provided in \cite{Caron-Huot:2018kta}. We briefly review some of the basic facts and then make the connection to the $\text{AdS}_2 \times \text{S}^2$ background. 

We begin with the simple observation that the $\text{AdS}_5 \times \text{S}^5$ is conformally equivalent to $10$-dimensional flat space, $\mathbb{R}^{1,9}$. However, the $\text{SO}(2,10)$ symmetry can be naturally interpreted as the conformal group in $\mathbb{R}^{1,9}$. The same statement can, also, be made for the $\text{AdS}_2 \times \text{S}^2$, with the $SO(2,4)$ symmetry being interpreted as the conformal group of the flat $\mathbb{R}^{1,3}$. 

Furthermore, the type IIB amplitude in flat-space is given by:
\begin{equation}\label{eq: iib_flat_ampl}
\mathcal{A} \sim G^{(10)}_N \delta^{16}(Q)\frac{1}{stu}
\,              .
\end{equation}
When we divide the flat-space amplitude given by \cref{eq: iib_flat_ampl} by the dimensionless effective coupling, $G^{(10)}_N \delta^{16}(Q)$, the stripped expression is annihilated by the generator of special conformal transformations:
\begin{equation}\label{eq: K_mu_def}
K_{\mu} = \sum_{i=1}^3\left(\frac{p_{i\mu}}{2}\frac{\partial}{\partial p_i^\nu}\frac{\partial}{\partial p_{i,\nu}}-p_i^\nu\frac{\partial}{\partial p_i^\nu}\frac{\partial}{\partial p_i^\mu}-\frac{d-2}{2}\frac{\partial}{\partial p_i^\mu}\right)
\,			.
\end{equation} 
Similarly, in $\text{AdS}_2 \times \text{S}^2$ the flat-space amplitude of hypermultiplets is given by
\begin{equation}\label{eq: flat_space_hypers}
\mathcal{A} = G_N \delta^4(Q) ~ \cdot 1
\,          .
\end{equation}
Upon dividing by the dimensionless combination $G_N \delta^4(Q)$ which is regarded as the effective coupling in this case, we obtain a stripped amplitude that is invariant under the action of \cref{eq: K_mu_def}. 

A final observation coming from \cite{Caron-Huot:2018kta} is that the form of the unmixed anomalous dimensions of double-trace operators concurs with the coefficients of the partial-wave decomposition of the $10$-dimensional amplitude. We do not investigate the relevant situation in this work, since the counterpart of this reasoning in $\text{AdS}_2 \times \text{S}^2$ was thoroughly scrutinized in \cite{Abl:2021mxo}.

Owing to the above similarities we proceed to extract the prediction of a hidden $\text{SO}(2,4)$ symmetry in the $\text{AdS}_2 \times \text{S}^2$. This hidden symmetry is the statement that the lowest-lying correlator, $H_{1111} = D_{1111}$, is promoted into a generating function. This is done by replacing the distances in AdS with distances in higher dimensions
\begin{equation} 
x^2_{ij} \rightarrow x^2_{ij}-t_{ij}
\,          .
\end{equation}
This is indicative of a $4$-dimensional conformal symmetry. Since the two sub-manifolds of $\text{AdS}_2 \times \text{S}^2$ are of equal radius, the background can be conformally mapped to the flat $R^{1,3}$ where $x^2_{ij}-t_{ij}$ is the conformally invariant distance. The generating function is:
\begin{equation}\label{eq: gen_fun_H}
\mathbf{H}(x_i,t_i) = H_{1111}(x^2_{ij}-t_{ij})
\,          .
\end{equation}
To obtain a correlator with general charges $H_{p_1 p_2 p_3 p_4}$, we only need perform a Taylor expansion of $\mathbf{H}(x_i,t_i)$ in powers of $t_{ij}$, and subsequently collect all the possible monomials $\prod\limits_{i<j} (t_{ij})^{\gamma_{ij}}$ that can appear in the correlator. There is only a finite number of such monomials for a given $H_{p_1 p_2 p_3 p_4}$. We provide some examples below:
\begin{equation}\label{eq: hidden_res}
\begin{aligned}
H_{11nn} \propto & ~ t^{n-1}_{34} D_{11nn} 
\,          ,
\\
H_{22nn} \propto & ~ \left(t_{12} t_{34} + (n-1) t_{13} t_{24} + (n-1)t_{14}t_{23} \right) D_{22nn} 
\,          ,
\\
H_{nnnn} \propto & ~ \left(t_{12} t_{34} + t_{13} t_{24} + t_{14} t_{23}\right)^{n-1}  D_{nnnn}
\,          .
\end{aligned}
\end{equation}
We can see that the results derived in \cref{eq: bootstrap_general_charges} agree with the prediction of hidden conformal symmetry given by \cref{eq: hidden_res} and one can check more examples explicitly.  
\section{Epilogue}\label{sec: epilogue}
In this work we have bootstrapped the $4$-point correlation function of hypermultiplets in $\text{AdS}_2 \times \text{S}^2$ supergravity. The approach we undertook relied only on crossing symmetry, the superconformal Ward identities and the bulk-point limit. Having explicitly derived the result for the $4$-point function, we proceeded to demonstrate that there is an exact agreement of our approach with the predictions of a hidden $4$-dimensional conformal symmetry. In this sense, we have provided a proof for the existence of this underlying structure in this simple setup. 

At the level of computing the holographic correlators, the take-home message is that hidden conformal symmetry is equivalent to imposing superconformal symmetry, the constraints of crossing symmetry and the consequences of the flat-space/bulk-point limit. 

The answer for the $4$-point correlator with arbitrary external weights in $\text{AdS}_2 \times \text{S}^2$  is given by: 
\begin{equation}
\langle\mathcal{O}_{p_1} \mathcal{O}_{p_2} \mathcal{O}_{p_3} \mathcal{O}_{p_4}\rangle \propto x^{w} \underbrace{\left(1- \frac{x}{y}\right)}_{\mathcal{R}} ~ \texttt{Wick}\big[\underbrace{t_1\ldots t_1}_{p_1 - 1} \ldots \underbrace{t_4\ldots t_4}_{p_4 - 1}\big] ~  \bar{D}_{p_1 p_2 p_3 p_4} 
\,          ,
\end{equation}
with $w$ being a positive number and the $t_i$ the null vectors on the $\text{S}^2$.

There are several fascinating avenues for future work: 

\begin{itemize}
\item It would be very interesting and useful to establish an appropriate formalism of Mellin amplitudes for this setup, perhaps along the lines of work of \cite{Bianchi:2021piu,Ferrero:2019luz}. The Mellin approach has been proved to be extremely useful in the higher-dimensional backgrounds, particularly in revealing hidden properties of holographic correlators. 

\item Owing to the simplicity of this setup, it would also be very desirable to extend the position-space bootstrap to higher-points, see the works \cite{Goncalves:2019znr,Alday:2022lkk,Goncalves:2023oyx,Alday:2023kfm} for recent progress in bootstrapping high-point correlators in different backgrounds. The simplicity of the answers in $\text{AdS}_2 \times \text{S}^2$ might be indicative of simplifications and suggestive for how hidden conformal symmetry works in the higher-dimensional cases that are not obvious directly in the higher-dimensional picture.    

\item We stress, once more, that the approach utilized here can be employed, in addition to some input from string theory, in the study of holographic defects when the co-dimension surface spans an $\text{AdS}_2$ subspace in the ambient geometry. This has already been exploited very successfully in \cite{Drukker:2020swu} for defects in the six-dimensional $(2,0)$ theory. Extending the logic to other theories, like the ABJM, should be straightforward. 
\end{itemize}

We hope to report to some of these aspects in the near future. 
\newpage
\section*{Acknowledgments} 
We are grateful to Xinan Zhou for suggesting this project to us, early collaboration, many useful discussions during the various stages of completion, and for reading a draft and offering his valuable comments. It is a pleasure to thank Hynek Paul and Michele Santagata for interesting comments and enlightening discussions. KCR is grateful to the National Technical University of Athens (NTUA) for the warm hospitality, where parts of this work were performed. The work of KCR is supported by starting funds from University of Chinese Academy of Sciences (UCAS), the Kavli Institute for Theoretical Sciences (KITS), and the Fundamental Research Funds for the Central Universities.
\newpage
\appendix
\section{Properties of \texorpdfstring{$D$}{D}-functions}\label{app: d.functions}
A $D$-function, denoted by $D_{\Delta_1 \ldots \Delta_n}$, represents a contact Witten diagram where the external operators have dimensions given by $\Delta_i$. Working in Euclidean $\text{AdS}_{d+1}$ with unit radius and in the Poincar\'e coordinates
\begin{equation}
ds^2 = \frac{dz^2_0 + d\vec{z} \cdot d\vec{z}}{z^2_0}
\,          ,
\end{equation}
this class of special functions is given by 
\begin{equation}\label{eq: funcD_def}
D_{\Delta_1 \ldots \Delta_n}(x_i)=\int \frac{d^d\vec{z}dz_0}{z_0^{d+1}}
\prod_{i=1}^n
\underbrace{\left(\frac{z_0}{z_0^2+(\vec{z}-\vec{x}_i)^2}\right)^{\Delta_i}}_{K_{\Delta_i}(z,x_i)}
\,          .
\end{equation}
where in the above $K_{\Delta_i}(z,x_i)$ is the bulk-to-boundary propagator.

These are $n$-point contact Witten diagrams in $\text{AdS}_{d+1}$ without derivatives. Note that we can represent contact diagrams with derivatives as $D$-functions, also, with shifted weights using that:
\begin{equation}\label{eq: contact_witten_ders}
\nabla^\mu K_{\Delta_1} \nabla_\mu K_{\Delta_2}=\Delta_1\Delta_2(K_{\Delta_1} K_{\Delta_2}-2x_{12}^2K_{\Delta_1+1} K_{\Delta_2+1}
)
\,			.
\end{equation}

It is very convenient to re-write to write the $D$-functions as functions of the conformal cross-ratios. This is achieved by extracting a kinematic factor. For the special case of $n=4$, $D$-functions can be written as $\bar{D}$-functions defined by:
\begin{equation}\label{eq: funcDbar_def}
\frac{ \prod\limits_{i=1}^4\Gamma(\Delta_i)}{\Gamma(\frac{1}{2}\Sigma_\Delta-\frac{1}{2}d)}\frac{2}{\pi^{\frac{d}{2}}}D_{\Delta_1\Delta_2\Delta_3\Delta_4}(x_i)=\frac{(x_{14}^2)^{\frac{1}{2}\Sigma_\Delta-\Delta_1-\Delta_4}(x^2_{34})^{\frac{1}{2}\Sigma_\Delta-\Delta_3-\Delta_4}}{(x^2_{13})^{\frac{1}{2}\Sigma_\Delta-\Delta_4}(x^2_{24})^{\Delta_2}}\bar{D}_{\Delta_1\Delta_2\Delta_3\Delta_4} (U,V)
\, 			,
\end{equation}
where we have used the shorthand $\Sigma_\Delta$ to denote the sum of the dimensions. 

Another particularly useful parameterisation of $D$-functions is provided by the use of the Feynman parameter. This leads to:
\begin{equation}\label{eq: feynman_param}
D_{\Delta_1\ldots \Delta_n}(x_i)=\frac{\pi^{\frac{d}{2}}\Gamma(\frac{1}{2}\Sigma_\Delta-\frac{1}{2}d)\Gamma(\frac{1}{2}\Sigma_\Delta)}{2\prod\limits_i\Gamma(\Delta_i)}\int \prod_j \frac{d\alpha}{\alpha_j}\alpha_j^{\Delta_j}\frac{\delta(\sum\limits_j\alpha_j-1)}{(\sum\limits_{k<l}\alpha_k\alpha_l x_{kl}^2)^{\frac{1}{2}\Sigma_\Delta}}
\,			.
\end{equation}

The reason that the the Feynman parameter representation is very useful is because it is clear from \cref{eq: feynman_param} that we have the following relations connecting $D$-functions with different weights
\begin{equation}\label{eq: derivative_weights}
D_{\Delta_1\ldots \Delta_i+1\ldots \Delta_j+1\ldots \Delta_n}(x_i)=\frac{d-\Sigma_\Delta}{2\Delta_i\Delta_j}\frac{\partial}{\partial x_{ij}^2}D_{\Delta_1\ldots \Delta_n}(x_i)
\,			.
\end{equation} 

For the special case of $n=4$ we can re-write the derivative relations, \cref{eq: derivative_weights}, in terms of the $\bar{D}$-functions:
\begin{equation}\label{eq: Dbars_opers}
\begin{split}
\bar{D}_{\Delta_1+1,\Delta_2+1,\Delta_3,\Delta_4}={}&-\partial_U \bar{D}_{\Delta_1,\Delta_2,\Delta_3,\Delta_4}
\,			,
\\
\bar{D}_{\Delta_1,\Delta_2,\Delta_3+1,\Delta_4+1}={}&(\Delta_3+\Delta_4-\tfrac{1}{2}\Sigma_\Delta-U\partial_U )\bar{D}_{\Delta_1,\Delta_2,\Delta_3,\Delta_4}
\,			,
\\
\bar{D}_{\Delta_1,\Delta_2+1,\Delta_3+1,\Delta_4}={}&-\partial_V \bar{D}_{\Delta_1,\Delta_2,\Delta_3,\Delta_4}
\,			,
\\
\bar{D}_{\Delta_1+1,\Delta_2,\Delta_3,\Delta_4+1}={}&(\Delta_1+\Delta_4-\tfrac{1}{2}\Sigma_\Delta-V\partial_V )\bar{D}_{\Delta_1,\Delta_2,\Delta_3,\Delta_4}
\,			,
\\
\bar{D}_{\Delta_1,\Delta_2+1,\Delta_3,\Delta_4+1}={}&(\Delta_2+U\partial_U+V\partial_V )\bar{D}_{\Delta_1,\Delta_2,\Delta_3,\Delta_4}
\,			,
\\
\bar{D}_{\Delta_1+1,\Delta_2,\Delta_3+1,\Delta_4}={}&(\tfrac{1}{2}\Sigma_\Delta-\Delta_4+U\partial_U+V\partial_V )\bar{D}_{\Delta_1,\Delta_2,\Delta_3,\Delta_4}
\,			.
\end{split}
\end{equation}

The simplest $n=4$ $\bar{D}$-function is given for $\{\Delta_i\}=1$. This results in the well-known scalar one-loop box integral in four dimensions which evaluates to \cite{Usyukina:1992jd}
\begin{equation}\label{eq: phi_1_def}
\bar{D}_{1111}\equiv\phi^{(1)}(z,\bar{z})=\frac{1}{z-\bar{z}}\left(2{\rm Li}_2(z)-2{\rm Li}_2(\bar{z})+\log(z\bar{z})\log\big(\frac{1-z}{1-\bar{z}}\big)\right)
\,			.
\end{equation}
The function $\phi^{(1)}$ obeys the differential recursion relations \cite{Eden:2000bk}:
\begin{equation}\label{eq: box_ders_recursion}
\begin{split}
\partial_z\phi^{(1)}={}&-\frac{\phi^{(1)}}{z-\bar{z}}-\frac{\log(-1+z)(-1+\bar{z})}{z(z-\bar{z})}+\frac{\log(z\bar{z})}{(-1+z)(z-\bar{z})}
\,			,
\\
\partial_{\bar{z}}\phi^{(1)}={}&\frac{\phi^{(1)}}{z-\bar{z}}+\frac{\log(-1+z)(-1+\bar{z})}{\bar{z}(z-\bar{z})}-\frac{\log(z\bar{z})}{(-1+\bar{z})(z-\bar{z})}
\,			.
\end{split}
\end{equation}

Using \cref{eq: box_ders_recursion} recursively allows us to write any $\bar{D}$-function as a linear combination of the basis functions $\{\phi^{(1)}(z,\bar{z}), \log U, \log V, 1\}$, with rational coefficients in $z$ and $\bar{z}$. 
\section{Some explicit \texorpdfstring{$\bar{D}$}{D}-functions in the holomorphic limit}\label{app: dbar.functions}
The holomorphic limit of the $\bar{D}$-functions necessary for the $\text{AdS}_2$ analysis considered in this work can be obtained in the following manner. We start by appropriately constructing the relevant following the prescription explained in \cref{app: d.functions} and then we carefully consider the limit $x \rightarrow \bar{x}$. For clarity and to facilitate the interested readers, we provide the expressions that are necessary in order to perform out position-space bootstrap for the case of the $\langle \mathcal{O}_1 \mathcal{O}_1 \mathcal{O}_1 \mathcal{O}_1 \rangle$ correlator.

For the $\langle \mathcal{O}_1 \mathcal{O}_1 \mathcal{O}_1 \mathcal{O}_1 \rangle$ correlator, we made an ansatz that contains all possible terms with $0$- and $2$- derivatives. The expressions for the necessary $\bar{D}$-functions are provided below
\begin{equation}\label{eq: explicitDbars1111}
\begin{aligned}
\bar{D}_{1111}			&=		\frac{2 x \log (x)+2(1- x) \log (1-x)}{(x-1) x}
\,		,
\\
\bar{D}_{1122}			&=		\frac{2 x^3 \log (x)+2 \left(-x^3+3 x-2\right) \log (1-x)+2 \left(x-x^2\right)}{6 (x-1)^2 x}
\,		,
\\
\bar{D}_{1212}			&=		\frac{ 2 \left(2 x^3-3 x^2\right) \log(x) + 2 \left(-2 x^3+3 x^2-1\right) \log (1-x) + 2 \left(x^2-x\right)}{6 (x-1)^2 x^2}
\,		,
\\
\bar{D}_{2112}			&=		\frac{2 \left(3 x^2-x^3\right) \log(x) + 2 \left(x^3-3 x^2+3 x-1\right) \log (1-x) + 2 \left(x^2-x\right)}{6 (x-1) x^2}
\,		,
\\
\bar{D}_{2121}			&=		\frac{2 \left(2 x^3-3 x^2\right) \log(x) + 2 \left(x^2-x\right)+2 \left(-2 x^3+3 x^2-1\right) \log (1-x)}{6 (x-1)^2 x^2}
\,		,
\\
\bar{D}_{2211}			&=		\frac{2 x^3 \log (x) + 2 \left(-x^3+3 x-2\right) \log (1-x) - 2 \left(x^2-x\right)}{6 (x-1)^2 x^3}
\,		,
\\
\bar{D}_{1212}			&=		\frac{2 \left(2 x^3-3 x^2\right) \log(x) +  2 \left(-2 x^3+3 x^2-1\right) \log (1-x) + 2 \left(x^2-x\right)}{6 (x-1)^2 x^2}
\,		.
\end{aligned}
\end{equation}

Of course we can construct the holomorphic limit of a $\bar{D}$-function with any charges in this way. We refrain, though, from providing more explicit expressions, as the formulae become quite lengthy.
\newpage
\bibliographystyle{ssg}
\bibliography{ads2s2}
\end{document}